\documentclass[lettersize,journal]{IEEEtran}
\usepackage{cite}
\usepackage{amsmath,amssymb,amsfonts}
\usepackage{algorithm}
\usepackage{algpseudocode}
\usepackage{graphicx}
\usepackage{textcomp}
\usepackage{tabularx}
\usepackage{array}
\usepackage{bbm}
\usepackage[hidelinks,colorlinks=true,linkcolor=blue,citecolor=blue,urlcolor=black]{hyperref}
\usepackage{academicons}
\usepackage[nolist]{acronym}
\usepackage{comment}
\usepackage{subcaption}
\usepackage{subfiles} 
\usepackage{textgreek}
\usepackage{listings}
\usepackage{balance}
\usepackage{soul}
\usepackage{multirow}
\usepackage[T1]{fontenc}
\usepackage{cuted} 
\usepackage{relsize}
\usepackage{booktabs}
\usepackage{longtable}
\usepackage{cleveref}
\usepackage[table]{xcolor}
\usepackage{makecell}
\usepackage{pifont}
\usepackage{caption}

\newcolumntype{Y}{>{\raggedright\arraybackslash}X}
\def\BibTeX{{\rm B\kern-.05em{\sc i\kern-.025em b}\kern-.08em
    T\kern-.1667em\lower.7ex\hbox{E}\kern-.125emX}}

\newsavebox{\mybox}

\begin{document}

\title{Constant-Envelope ISAC via FM-OFDM: Analytical Framework and Receiver Design}

\author{Amir Bouziane,~\IEEEmembership{}
        and~H\"{u}seyin~Arslan,~\IEEEmembership{Fellow,~IEEE}
\thanks{Amir Bouziane is with Electrical and Electronics Engineering, School of Engineering and Natural Sciences, Istanbul Medipol University, 34810 Istanbul, T\"{u}rkiye (e-mail: \href{bouziane.amir@std.medipol.edu.tr}{bouziane.amir@std.medipol.edu.tr}).}
\thanks{H\"{u}seyin Arslan is with Electrical and Electronics Engineering, School of Engineering and Natural Sciences, Istanbul Medipol University, 34810 Istanbul, T\"{u}rkiye.}
\thanks{This work has been submitted for publication.}
\thanks{CORRESPONDING AUTHOR: AMIR BOUZIANE (e-mail: \href{bouziane.amir@std.medipol.edu.tr}{bouziane.amir@std.medipol.edu.tr}).}
}

\markboth{Constant-Envelope ISAC via FM-OFDM: Analytical Framework and Receiver Design}{A.BOUZIANE \textit{et al.}}


\maketitle

\begin{acronym}
  \acro{3D}{three-dimensional}
  \acro{3GPP}{3rd Generation Partnership Project}
  \acro{4G}{fourth generation}
  \acro{5G}{fifth generation}
  \acro{5G-NR}{5G New Radio}
  \acro{6G}{sixth generation}
  \acro{OTFS}{orthogonal time frequency space}
  \acro{AAoA}{azimuth angle of arrival}
  \acro{AAoD}{azimuth angle of departure}
  \acro{A2A}{air-to-air}
  \acro{A2G}{air-to-ground}
  \acro{ACI}{adjacent-channel interference}
  \acro{RMS}{Root Mean Square}
  \acro{ACLR}{Adjacent Channel Leakage Ratio}
  \acro{AWGN}{additive white Gaussian noise}
  \acro{BER}{bit error rate}
  \acro{BPSK}{binary phase-shift keying}
  \acro{BS}{base station}

  \acro{NR-V2X}{New Radio Vehicle-to-Everything}
  \acro{KPI}{key parameter indicator}
  \acro{SDR}{software-defined radio}
  \acro{DVB}{digital video broadcasting}
  \acro{DPD}{digital predistortion}
  \acro{OOB}{Out-of-Band}
  \acro{BW}{bandwidth}
  \acro{PLL}{phase-locked loop}
  \acro{AF}{ambiguity function}
  \acro{PHY}{physical-layer}
  \acro{SI}{self-interference}
  \acro{LUT}{Look-up Table}
  \acro{RF}{radio frequency}

  \acro{CCD}{communication-centric design}
  \acro{CD-NOMA}{code-domain non-orthogonal multiple access}
  \acro{CE}{constant envelope}
  \acro{CE-OFDM}{constant-envelope orthogonal frequency-division multiplexing}
  \acro{FM-OFDM}{frequency modulated orthogonal frequency-division multiplexing}
  \acro{CFO}{carrier frequency offset}
  \acro{CCI}{co-channel interference}
  \acro{CP}{cyclic prefix}
  \acro{CP-OFDM}{cyclic-prefix orthogonal frequency-division multiplexing}
  \acro{CRLB}{Cramér Rao lower bound}
  \acro{CSI}{channel state information}
  \acro{PN}{phase noise}
  \acro{DAC}{digital-to-analog converter}
  \acro{DC}{direct current}
  \acro{DFRC}{dual-functional radar communication}
  \acro{DFT}{discrete Fourier transform}
  \acro{DL}{downlink}

  \acro{EAoA}{elevation angle of arrival}
  \acro{EAoD}{elevation angle of departure}
  \acro{EVD}{eigenvalue decomposition}

  \acro{FFT}{fast Fourier transform}
  \acro{FMCW}{frequency-modulated continuous wave}
  \acro{FoV}{field of view}
  \acro{FR}{frequency range}

  \acro{G2A}{ground-to-air}
  \acro{G2G}{ground-to-ground}
  \acro{GEO}{geostationary equatorial orbit}

  \acro{HAP}{high-altitude platform}
  \acro{HIBS}{HAP-based IMT base stations}
  \acro{HRLLC}{hyper-reliable and low-latency communication}

  \acro{IBI}{inter-beam interference}
  \acro{ICI}{inter-carrier interference}
  \acro{IDFT}{inverse discrete Fourier transform}
  \acro{IFFT}{inverse fast Fourier transform}
  \acro{IoT}{Internet of Things}
  \acro{ISAC}{Integrated sensing and communication}
  \acro{ISI}{inter-symbol interference}
    \acro{CRB}{ cramér–rao bound}
  \acro{JD}{joint design}
  \acro{JSAC}{joint sensing and communication}

  \acro{LEO}{low Earth orbit}
  \acro{LLR}{log-likelihood ratio}
  \acro{LOS}{line of sight}
  \acro{LTE}{Long Term Evolution}

  \acro{MEO}{medium Earth orbit}
  \acro{MIMO}{multiple-input multiple-output}
  \acro{ML}{maximum-likelihood}
  \acro{mmWave}{millimeter-wave}
  \acro{MPA}{message passing algorithm}
  \acro{MSE}{mean squared error}
  \acro{MUSIC}{multiple signal classification}

  \acro{NMSE}{normalized mean squared error}
  \acro{NOMA}{non-orthogonal multiple access}
  \acro{NTN}{non-terrestrial network}

  \acro{OFDM}{orthogonal frequency-division multiplexing}
  \acro{OFDM-IM}{OFDM with index modulation}
  \acro{OFDM-DM}{OFDM with directional modulation}
  \acro{PA}{power amplifier}
  \acro{PAPR}{peak-to-average power ratio}
  \acro{PDF}{probability density function}
  \acro{PDMA}{power-domain multiple access}
  \acro{PD-NOMA}{power-domain non-orthogonal multiple access}
  \acro{PMCW}{phase-modulated continuous wave}

  \acro{QAM}{quadrature amplitude modulation}
  \acro{QPSK}{Quadrature Phase Shift Keying }
  \acro{QoS}{quality of service}
  \acro{MF}{matched filter}
  \acro{RCS}{radar cross section}
  \acro{RDM}{range Doppler map}
  \acro{RF}{radio frequency}
  \acro{RIS}{reconfigurable intelligent surface}
  \acro{RMSE}{root mean square error}
  \acro{RSSI}{received signal strength indicator}
  \acro{Rx}{receiver}

  \acro{SCMA}{sparse code multiple access}
  \acro{SIC}{successive interference cancellation}
  \acro{SINR}{signal-to-interference-plus-noise ratio}
  \acro{SNR}{signal-to-noise ratio}
  \acro{SVD}{singular value decomposition}

  \acro{Tbps}{terabits per second}
  \acro{TDD}{time-division duplex}
  \acro{THz}{terahertz}
  \acro{TN}{terrestrial network}
  \acro{TR}{Technical Report}
  \acro{TS}{Technical Specification}
  \acro{Tx}{transmitter}

  \acro{UAV}{unmanned aerial vehicle}
  \acro{UE}{user equipment}
  \acro{UL}{uplink}
  \acro{URPA}{uniform rectangular planar array}

  \acro{V2I}{vehicle-to-infrastructure}
  \acro{V2X}{vehicle-to-everything}

  \acro{XR}{extended reality}
\end{acronym}

\begin{abstract}
Integrated Sensing and Communication (ISAC) systems face stringent hardware constraints, particularly regarding the high Peak-to-Average Power Ratio (PAPR) of standard OFDM, which necessitates power amplifier (PA) back-off and reduces sensing range. This paper investigates Frequency-Modulated OFDM (FM-OFDM) as a constant-envelope solution capable of operating in the PA saturation region, thereby maximizing output power without the non-linear distortion penalties typical of conventional waveforms. We derive a comprehensive analytical framework for FM-OFDM in doubly dispersive channels, explicitly quantifying the inter-carrier interference (ICI) dynamics and effective channel gains in the discriminator domain. To address the unique phase structure of the waveform, we propose a tailored sensing receiver architecture utilizing slow-time phase differencing for robust velocity estimation. Unlike prior works, we evaluate performance under a strictly normalized bandwidth constraint ($B_{99}$), ensuring a fair comparison against CP-OFDM and Constant-Envelope OFDM (CE-OFDM). Simulation results demonstrate that FM-OFDM maintains superior detection accuracy and low BER even under fully saturated PA conditions and high Doppler shifts, validating its suitability for hardware-constrained ISAC transceivers.
\end{abstract}

\begin{IEEEkeywords}
6G, ISAC, Constant Envelope, FM-OFDM, CE-OFDM, Cp-OFDM, Radar, Waveform Design, Doubly Dispersive Channels.
\end{IEEEkeywords}

\section{INTRODUCTION}
\IEEEPARstart{T}{he} upcoming \ac{6G} wireless networks promise a transformation of global connectivity, targeting ultra-high data rates up to terabits per second (Tbps), ultra-low latency (sub-0.1 ms), massive connectivity density (millions of devices per km$^2$), and sustainable energy usage \cite{5GAmericas2024_3GPPTrends}. These capabilities are fundamental for supporting next-generation applications such as immersive holography, real-time digital twins, remote telesurgery, autonomous mobility systems, and massive-scale intelligent infrastructure \cite{9144301}. To meet these extreme performance requirements, a fundamental redesign of physical-layer protocols, signal processing, and spectrum usage is necessary \cite{ITU_M2160_2023,3GPP_RAN_R19_ISAC}.

One of the promising upcoming enabling this evolution is \ac{ISAC}, where the radio interface is designed to simultaneously support data exchange and environmental perception. Unlike traditional systems that treat sensing and communication as separate tasks, often isolating them via orthogonal resources like time, frequency, or antennas. \ac{ISAC} enables both functionalities to coexist on the same waveform, hardware, and spectrum. This shared architecture reduces hardware redundancy, increases spectral efficiency, improves latency, provides situational, and interference management \cite{liu2025uncovering,liu2022survey,fang2022joint}.

Recent standardization efforts highlight the importance of \ac{ISAC}. IEEE 802.11bf introduces protocols for Wi-Fi-based sensing, enabling gesture detection, respiration monitoring, and device free localization \cite{IEEE80211bf_2025}.
\ac{3GPP} has incorporated \ac{ISAC} primitives into Releases 18 and 19, supporting sidelink positioning, radar aided beamforming, and joint vehicular sensing in \ac{NR-V2X} scenarios. On the other side, ETSI’s Industry Specification Group on \ac{ISAC} is actively defining requirements, \ac{KPI}s, and interoperability frameworks to facilitate deployment in industrial automation and smart transportation networks \cite{ETSI_GR_ISAC_001,liu2024next}. 
All throughout wireless communication history radar and communication systems evolved along separate paths: radar prioritized precise range and velocity estimation using chirps and pulse Doppler techniques, while communication systems emphasized spectral efficiency and capacity. \\
This convergence demands that the waveform design community revisit metrics such as ambiguity functions, \ac{CRB}, spectral efficiency, and \ac{PAPR} trade-offs \cite{liao2025pulse,10685511}. The critical enabler of \ac{ISAC} is waveform design. For \ac{ISAC} to be effective, the waveform must simultaneously meet multiple requirements. These include time frequency resolution for accurate sensing, low \ac{BER} and robustness for reliable communication, low \ac{PAPR} to ensure power efficiency, and flexible spectral occupation to comply with regulatory constraints \cite{sturm2011waveform}. Existing waveform-design strategies for \ac{ISAC} generally fall into three main categories. \\
The first category, sensing-centric designs, includes waveforms such as \ac{PMCW} and \ac{FMCW}. Optimized for sensing and provide excellent range Doppler performance, however, they often perform poorly in terms of data transmission efficiency. Second category is communication centric designs. These use traditional communication waveforms such as \ac{OFDM} or single carrier systems. While such waveforms are effective for data transmission, they suffer from limitations in sensing resolution and exhibit high \ac{PAPR}, which reduces power efficiency. The most promising category comprises joint designs. Which aims to optimize sensing and communication within a unified waveform framework. Examples include \ac{CE-OFDM}, \ac{DFT}-spread-\ac{OFDM}, and other hybrid strategies designed to balance dual-function performance \cite{9924202,liu2022integrated}.

\ac{OFDM}, as used in \ac{LTE}, \ac{5G-NR}, and Wi-Fi, remains the dominant communication waveform due to its flexible multicarrier structure and efficient frequency domain equalization. However, \ac{OFDM} suffers from a critical drawback: its inherently high \ac{PAPR} requires that the \ac{PA} operate in its nonlinear region. This inefficiency is particularly problematic in mmWave and THz bands where hardware linearity and thermal constraints limit achievable output power \cite{wei2023integrated}.

To address these limitations, constant-envelope (CE) waveforms have gained significant attention. \ac{CE-OFDM} modulates the phase of \ac{OFDM} signals to achieve 0 dB \ac{PAPR}, permitting fully saturated \ac{PA} operation. However, \ac{CE-OFDM} is sensitive to phase-unwrapping errors, especially under low \ac{SNR}, which hinders accurate Doppler estimation \cite{chung1999constant,thompson2008constant,felton2023gradient}. A promising alternative is \ac{FM-OFDM} \cite{hernando2022frequency}, which employs instantaneous frequency modulation of a baseband \ac{OFDM} signal. By introducing a tunable modulation index $m$, \ac{FM-OFDM} offers a flexible trade off between communication quality and sensing precision.

A unified analytical framework for \ac{FM-OFDM} in \ac{ISAC} scenarios remains open for research. While foundational work in \cite{hernando2022frequency,hernando2025channel} established \ac{FM-OFDM}'s robustness for high-mobility communications, these studies did not address specific radar challenges.
For practical \ac{ISAC} deployment, several key challenges must be addressed such as the spectral trade off \cite{wang2022triangular}, target seperability resolution, \cite{GonzalezPrelcic_ProcIEEE_2024,Oliveira_PN_OFDM_ISAC_2024}, and hardware limitations such as phase noise and \ac{CFO} in the discriminator path \cite{Chen_Golay_CE_OFDM_ISAC_2024}.

Under the light of what is discussed above, our motivation is an \ac{ISAC}-specific, discriminator domain treatment of \ac{FM-OFDM} that addresses these issues. We focus on the comparative analysis of \ac{FM-OFDM} against\ac{CP-OFDM} and \ac{CE-OFDM}, to establish its viability under common spectral constraints between these waveforms.

\subsection{CONTRIBUTIONS}
This work is an end to end assesement of \ac{FM-OFDM} for \ac{ISAC}, distinguishing it from prior communication centric studies by developing a dedicated sensing architecture and enforcing  spectral constraints. The main contributions are summarized as follows:
\begin{itemize}
\item[i.] We derive the complete input-output relationship for \ac{FM-OFDM} in doubly dispersive channels. We explicitly quantify the sensing-specific \ac{ICI} and effective channel gains.
\item[ii.]  We propose a discriminator domain sensing receiver. To overcome the failure of standard Doppler estimation caused by \ac{FM-OFDM}'s deterministic phase variations, we use a slow time phase-differencing estimator that extracts velocity information without requiring complex phase synchronization under the assumption of no phase unwrapping.
\item[iii.]  We conduct a strictly fair comparative evaluation against \ac{CP-OFDM} and \ac{CE-OFDM}. Unlike previous works, we normalize the occupied bandwidth ($B_{99}$) across all waveforms to decouple the resolution gains of high-$m$ \ac{FM-OFDM} from simple spectral expansion.
\item[iv.] We systematically analyze the trade-offs between the modulation index $m$, spectral efficiency, and estimation accuracy. Simulations demonstrate that under fair spectral constraints, \ac{FM-OFDM} achieves range velocity resolution robust to Doppler effects in high mobility regimes while maintaining the power efficiency benefits of a constant envelope.
\end{itemize}

\subsection{PAPER ORGANIZATION}
The remainder of this article is organized as follows. Section \ref{sec:fmofdm_io_main} derives the \ac{FM-OFDM} input–output relationship over doubly dispersive channels. Section \ref{sec:mod_index} analyzes the modulation index trade-offs. Section \ref{sec:radar_processing} develops the sensing receiver, discriminator-domain matched-filter range compression, and slow-time phase-difference Doppler estimation. Section \ref{sec:sim_results} presents communication and sensing evaluations (\ac{BER}, range/velocity \ac{RMSE}, and range-Doppler maps) and studies the role of the modulation index $m$. Finally, Section \ref{sec:conclusion} concludes the paper, and Appendix \ref{app:weights} provides supporting derivations.


\section{FM-OFDM INPUT–OUTPUT RELATIONSHIP}
\label{sec:fmofdm_io_main}

\begin{figure}[t]
    \centering
    \includegraphics[width=\columnwidth]{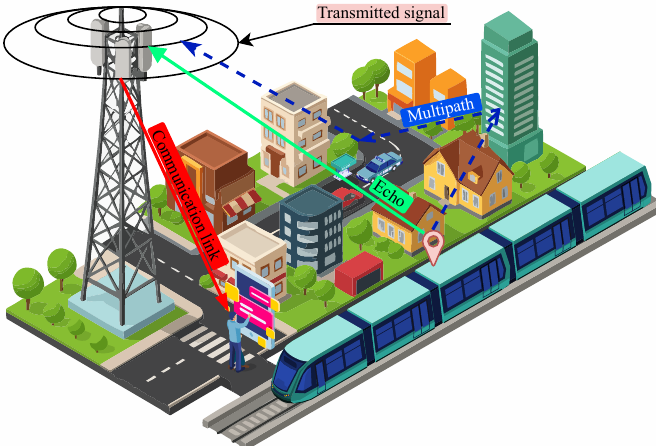}
    \caption{FM-OFDM transceiver with a communication user and a mobile target.}
    \label{fig:scenario1}
\end{figure}

\textcolor{black}{We consider a base station that simultaneously performs wireless communication and environmental sensing using a unified waveform, as shown in Fig. \ref{fig:scenario1}. In this monostatic \ac{ISAC} configuration, the base station functions as both the transmitter and the radar receiver for sensing. It delivers data to communication users and probes the environment via reflected echoes from surrounding targets. A fundamental challenge in such a monostatic setup is the \ac{SI} that arises from the powerful transmitted signal leaking into the co-located sensing receiver \cite{osorio2025rise}. However, it is common to achieve a high degree of isolation through physical separation of the transmit and receive antennas, antenna directionality, and the use of circulators \cite{chen2022antenna}. Stand-alone monostatic \ac{ISAC} systems are considered a practical option for initial deployments because they simplify system design and synchronization requirements \cite{jeong2025interference}. Therefore, consistent with other works that focus on waveform performance \cite{zhuo2022multibeam}, our analysis proceeds under the assumption that sufficient passive isolation and other self-interference techniques are in place to suppress this leakage to a level below the receiver's noise floor. This allows us to focus on the waveform's performance against external channel effects. The system transmits a CE \ac{FM-OFDM} waveform.}
\\

\textcolor{black}{While the fundamental signal generation and receiver structure are based on the principles established for \ac{FM-OFDM} communication systems \cite{hernando2022frequency,hernando2025channel}, the novel contribution of this section is the rigorous derivation of the complete input-output relationship and the effective channel matrix for a doubly dispersive \ac{ISAC} scenario. This analytical framework is essential for the subsequent sensing performance evaluation.}

\subsection{Signal Model and Transmit Waveform}
We adopt the CE \ac{FM-OFDM} transmit structure established in \cite{hernando2022frequency}. Let $x[n]$ denote the real-valued, normalized \ac{OFDM} baseband signal. The transmit phase is generated as $\phi[n] = \phi_0 + 2\pi T_s \sum_{u=0}^{n} m f_\Delta x[u]$, yielding the CE signal $s[n] = A e^{j\phi[n]}$. 
While the transmit generation is standard, the interaction of this waveform with a doubly dispersive sensing channel requires a novel characterization of the interference terms, which we derive below.
\subsection{CHANNEL MODEL}
Over one $N$-sample block we use a blockwise-circular, time-varying multipath baseband model with $P$ components. The discrete-time channel impulse response at $n$ and delay tap $\ell$ is
\begin{align}
    h[n,\ell] \triangleq \sum_{p=0}^{P-1} a_p\,e^{j2\pi \nu_p nT_s}\,\delta[\ell-\ell_p],
\end{align}
\textcolor{black}{where $a_p\in\mathbb{C}$ is the complex path gain, $\nu_p$ is the Doppler shift in Hz, $\ell_p\in\{0,\ldots,N{-}1\}$ is the integer sample delay, and $T_s$ is the sampling interval. The \ac{OFDM} symbol duration is $T_{\mathrm{sym}}=N T_s$.} The corresponding received complex baseband sample is
\begin{align}
    r[n] \triangleq \sum_{p=0}^{P-1} a_p\,e^{j2\pi \nu_p (n-\ell_p)T_s}\,s[n-\ell_p] \;+\; w[n].
    \label{eq:rx}
\end{align}
Here $a_p\in\mathbb{C}$ is the complex path gain of component $p$, $\nu_p$ is its Doppler shift, $\ell_p\in\{0,\ldots,N{-}1\}$ is the integer sample delay under the circular model, and $w[n]\sim\mathcal{CN}(0,\sigma^2)$ is complex \ac{AWGN}. We reference delays relative to a dominant path $\ell_0$ via $\Delta\ell_p\triangleq \ell_p{-}\ell_0$, and denote the \ac{OFDM}-symbol duration by $T_{\mathrm{sym}}=N T_s$. A constant \ac{CFO} can be absorbed into the Doppler terms $\nu_p$ and will be approximately removed later by block de-meaning.

\subsection{FM DEMODULATION AND OFDM PROCESSING}
\label{sec:fm_demod}
We consider a limiter–discriminator receiver: a hard limiter removes amplitude, followed by a frequency discriminator with gain $K_V$, \textcolor{black}{which is a scaling factor that converts phase differences to frequency estimates. In our analysis, we normalize $K_V = 1$ without loss of generality, as it appears as a common scaling factor in all expressions and cancels in relative performance comparisons.} Define the limited signal
\begin{equation}
    z[n]\triangleq \frac{r[n]}{|r[n]|}\in\mathbb{C},\qquad n=0,\ldots,N-1.
\end{equation}
The one-sample phase-difference discriminator produces the instantaneous-frequency estimate \cite{hernando2022frequency}
\begin{equation}
    \widehat f[n] \;=\; \frac{1}{2\pi T_s}\,\angle\!\big(z[n]\,z^*[n-1]\big), \label{eq:disc}
\end{equation}
and the discriminator output is
\begin{equation}
    y[n] \;=\; K_V\,\widehat f[n] \;=\; K_V\,\frac{\angle(z[n]z^*[n-1])}{2\pi T_s}.
\end{equation}
Using the small–phase-increment approximation together with the phasor-sum property of instantaneous frequency \cite{loughlin2002comments}, and under high-\ac{SNR} with mild within-block time selectivity, the limiter–discriminator output admits the following weighted-sum representation (derivation in Appendix\ref{app:weights}):
\begin{equation}
    y[n] \;=\; K_V \sum_{p=0}^{P-1} \beta_p[n]\Big(\nu_p + m f_\Delta\,x[n-\ell_p]\Big) + v[n],
    \label{eq:demod_sum}
\end{equation}
where $\beta_p[n]\!\ge\!0$ and $\sum_p \beta_p[n]\!=\!1$ (slowly varying weights determined by the post-limiter phasor geometry), $\nu_p$ is the Doppler shift , $x[\cdot]$ is the normalized \ac{OFDM} baseband signal, $m f_\Delta$ is the frequency deviation scale, $K_V$ is the discriminator gain, and $v[n]$ models discriminator noise (distinct from the front-end noise $w[n]$ in \eqref{eq:rx}). Aligning to a dominant delay $\ell_0$ and removing the block mean approximately suppress a constant \ac{CFO}:
\begin{align}
    y_{\rm s}[n]      &\triangleq y[n+\ell_0], \nonumber\\
    \bar y_{\rm s}[n] &\triangleq y_{\rm s}[n]-\frac{1}{N}\sum_{u=0}^{N-1}y_{\rm s}[u]. \label{eq:meanrm}
\end{align}
\textcolor{black}{Within a single $N$-sample \ac{DFT} block, we approximate $\beta_p[n]\approx\beta_p$ as blockwise constant under the slow variation conditions discussed in Section \ref{subsec:slow_variation}. The weights $\beta_p$ satisfy $\beta_p\propto |a_p|$ with $\sum_p\beta_p=1$, representing the normalized contribution of each path to the composite instantaneous frequency. Physically, these weights emerge from the vector sum of complex phasors at the limiter output: stronger paths and those with constructive phase alignment contribute more heavily to the instantaneous frequency estimate, while weaker or destructively interfering paths contribute less.}
\textcolor{black}{\subsubsection{Slow Variation Assumption and Weight Dynamics}
\label{subsec:slow_variation}
The derivation of the effective channel matrix relies on the approximation $\beta_p[n] \approx \beta_p$, where the time-varying weights are treated as constant over an OFDM symbol duration. This assumption holds under practical channel conditions where:
The maximum Doppler spread $f_D^{\text{max}}$ satisfies $f_D^{\text{max}} T_{\text{sym}} \ll 1$, the channel coherence time is much longer than the symbol duration $T_{\text{sym}}$ and the relative phases between multipath components change slowly.\\
These conditions are typical in most mobile scenarios below 6 GHz and moderate velocities. The weights $\beta_p$ represent the normalized contribution of each path to the instantaneous frequency estimate after limiter processing. Physically, they capture the relative dominance of each path in the constructive/destructive interference at the limiter output, determined by both path amplitudes $|a_p|$ and their relative phases.
When these conditions are violated (e.g., in extremely high mobility or rapidly time-varying channels), the deviations $\delta\beta_p[n] = \beta_p[n] - \beta_p$ become significant and induce the \ac{ICI} terms analyzed in Eq. \eqref{eq:ICI_kernel}.}

\subsubsection{CFO MITIGATION VIA DE-MEANING}
By adding a common \ac{CFO} \(\nu_0\) to the per path Dopplers for every \(p\). The limiter–discriminator output for a single \(N\)-sample useful block is represented by \(y[n]\). We assume that blockwise-slow variation model \(\beta_p[n]\approx \beta_p\) within the block. Next, with reference delay \(\ell_0\) aligned to the delay-aligned sequence \(y_s[n]\triangleq y[n+\ell_0]\) \cite{haykin2008communication}, the block mean contains \(K_V\nu_0\), which is removed by de-meaning; residual drift yields a small bias \(O\!\big(\sum_p \beta_p\,\dot\nu_p\,T_{\mathrm{sym}}\big)\).

Letting $\Delta\ell_p\triangleq \ell_p-\ell_0$, taking the $N$-point \ac{DFT} of the de-meaned sequence, and recalling $X[q]\triangleq \sum_{n=0}^{N-1}x[n]e^{-j\frac{2\pi q n}{N}}$, yields
\begin{align}
    Y[k] & \triangleq \sum_{n=0}^{N-1} \bar y_{\rm s}[n] e^{-j\frac{2\pi kn}{N}} \nonumber                                                     \\
        & \approx K_V m f_\Delta\sum_{p}\beta_p
    \sum_{n=0}^{N-1} x[n-\Delta\ell_p] e^{-j\frac{2\pi kn}{N}} \nonumber                                                                                  \\
        & = K_V m f_\Delta\sum_{p}\beta_p \sum_{q=0}^{N-1} X[q] e^{-j\frac{2\pi q\Delta\ell_p}{N}}
    \sum_{n=0}^{N-1} e^{j\frac{2\pi (q-k)n}{N}} \nonumber                                                                                                 \\
        & = \underbrace{K_V m f_\Delta\Big(\sum_{p}\beta_p e^{-j\frac{2\pi k\Delta\ell_p}{N}}\Big)}_{H^{\rm (diag)}_{k}} X[k] + W_k.
    \label{eq:Yk_ideal}
\end{align}
where $W[k]$ is the DFT of the residual discriminator noise and de-meaning remainder. Thus, after block de-meaning, the effective channel is diagonal in subcarrier index $k$ with a delay-induced linear phase across subcarriers.
\textcolor{black}{The key step in this derivation is the orthogonality property of the DFT basis functions. The inner sum over $n$ evaluates to:
\begin{align}
    \sum_{n=0}^{N-1} e^{j\frac{2\pi (q-k)n}{N}} = 
    \begin{cases}
        N & \text{if } q = k \\
        0 & \text{if } q \neq k
    \end{cases}
    = N\delta[q-k]
\end{align}
This orthogonality causes all cross-terms ($q \neq k$) to vanish, leaving only the diagonal components where $q = k$. The resulting expression shows that after block de-meaning, the effective channel becomes diagonal in the subcarrier domain with phase shifts induced by the multipath delays.
}\\
\textcolor{black}{When the slow variation assumption is violated, the time-varying components $\delta\beta_p[n]$ induce \ac{ICI}. The \ac{ICI} kernel can be understood as a two-dimensional leakage function: the inner sum over $n$ represents the \ac{DFT} of the weight variations $\delta\beta_p[n]$, while the outer sum over $q$ captures how this spectral leakage spreads energy from subcarrier $q$ to subcarrier $k$.}
When $\beta_p[n]$ varies within the block, we express it as $\beta_p[n]=\beta_p+\delta\beta_p[n]$. This perturbation induces off-diagonal terms:
\begin{equation}
    \begin{aligned}
        Y_k^{\rm (ICI)}= K_V m f_\Delta \ \ \ \ \ \ \ \ \ \ \ \ \ \ \ \ \ \ \ \ \ \ \ \ \ \ \ \ \ \ \ \ \ \ \ \ \ \ \ \ \ \ \ \ \ \ \ \\ 
        \times\sum_{q=0}^{N-1}\!\Bigg(\sum_p \!\sum_{n=0}^{N-1}\! \delta\beta_p[n]\,
        \frac{e^{j\frac{2\pi q (n-\Delta\ell_p)}{N}}}{N}\,e^{-j\frac{2\pi kn}{N}}\!\Bigg)       \times X[q],
    \end{aligned}
    \label{eq:ICI_kernel}
\end{equation}
which vanish when $\delta\beta_p[n]\equiv 0$. See \cite{boashash2002estimating,flandrin1998time} for background on instantaneous frequency of multicomponent signals and the amplitude-weighted interpretation that motivates the slowly varying weights $\beta_p[n]$.
 
\subsection{EFFECTIVE CHANNEL MATRIX}
Stacking  
\begin{align*}
    \mathbf{Y}\triangleq[Y[0],\ldots,Y[N{-}1]]^\top, & \\
    \ \ \ \ \ \ \mathbf{X} \triangleq[X[0],\ldots,X[N{-}1]]^\top, &\\
    \mathbf{W}\triangleq[W[0],\ldots,W[N{-}1]]^\top,
\end{align*}
Then
\begin{equation}
    \mathbf{Y} \;=\; \mathbf{H}_{\rm eff}\,\mathbf{X} \;+\; \mathbf{W},
    \label{eq:io}
\end{equation}
with matrix entries
\begin{equation}
    H^{\rm eff}_{k,r}
    \;=\; K_V\,m f_\Delta \!\left(\sum_{p} \beta_p\,e^{-j\frac{2\pi}{N}k(\ell_p-\ell_0)}\right)\!\delta[k-r]
           \;+\; H^{\rm (ICI)}_{k,r},
    \label{eq:Heff}
\end{equation}
so the per-subcarrier relation is
\begin{equation}
    Y[k] \;=\; H^{\rm eff}_{k,k}\,X[k] \;+\; \sum_{r\neq k} H^{\rm eff}_{k,r}\,X[r] \;+\; W[k].
    \label{eq:Yk_final}
\end{equation}
Here $\Delta\ell_p\triangleq \ell_p-\ell_0$ and $H^{\rm (ICI)}_{k,r}$ captures the \ac{ICI} induced by within-block weight variations $\delta\beta_p[n]$ \eqref{eq:ICI_kernel}. For a static single-path channel with $\Delta\ell_0=0$, the matrix is strictly diagonal with $H^{\rm eff}_{k,r}=K_V\,m f_\Delta\,\delta[k-r]$, i.e., $Y[k]=K_V\,m f_\Delta\,X[k]+W[k]$.

For fixed discriminator noise, both the diagonal gain and any residual \ac{ICI} scale linearly with $m$. For unit-variance $x[n]$, a Carson-like occupied-bandwidth estimate is
\begin{equation*}
     B_{99}\;\approx\;2\big(B_x+\eta\,m f_\Delta\big),
\end{equation*}
where $B_x$ is the occupied baseband of $x[n]$ and $\eta\in[1,2]$ depends on the distribution of $x[n]$ \cite{haykin2008communication}. Together with the aliasing constraint \eqref{eq:alias}, this yields the usual trade-off: increasing $m$ improves post-demodulation \ac{SNR} via larger frequency swings but increases occupied bandwidth and aliasing risk; decreasing $m$ is spectrally compact but less robust to discriminator noise.

\subsubsection{CFO-FREE EFFECTIVE CHANNEL}
Let $\bar{y}_{\rm s}[n]$ be the de-meaned, delay-aligned discriminator output in \eqref{eq:meanrm}. Define the $N$-point analysis \ac{DFT}s
\begin{equation*}
    \bar{Y}[k] \triangleq \sum_{n=0}^{N-1} \bar{y}_{\rm s}[n]\,e^{-j\frac{2\pi}{N}kn},\qquad
    X[r] \triangleq \sum_{n=0}^{N-1} x[n]\,e^{-j\frac{2\pi}{N}rn}.
\end{equation*}
Then
\begin{equation}
\bar{Y}[k] \;=\; \sum_{r=0}^{N-1} H_{\mathrm{eff}}[k,r]\,X[r] \;+\; W[k],
\end{equation}
with a diagonal “static” term
\begin{equation}
        H_{\mathrm{diag}}[k,r] \;=\; K_V\,m f_\Delta\!\left(\sum_{p} \beta_p\,e^{-j\frac{2\pi}{N}k(\ell_p-\ell_0)}\right)\!\delta[k-r],
\end{equation}
and an \ac{ICI} term driven by within-block weight variations $\delta\beta_p[n]\triangleq \beta_p[n]-\beta_p$:
\begin{align}
    H_{\mathrm{ICI}}[k,r]  
     =  K_V\,m f_\Delta \sum_{p}\Bigg(\frac{1}{N}\sum_{n=0}^{N-1}\delta\beta_p[n]\,
        e^{-j\frac{2\pi}{N}(k-r)n}\Bigg)\,\\ \times  e^{-j\frac{2\pi}{N}r(\ell_p-\ell_0)}.
\end{align}
Thus $H_{\mathrm{eff}}[k,r]=H_{\mathrm{diag}}[k,r]+H_{\mathrm{ICI}}[k,r]$, and no explicit \ac{CFO} term appears in $H_{\mathrm{eff}}$ because the block de-meaning removed the constant component in time. For a static channel ($\nu_p{=}0$), the per-block mean in \eqref{eq:meanrm} is already zero and $\mathbf{H}_{\rm eff}$ is strictly diagonal with

\begin{equation}
    H^{\rm eff}_{k,k} \;=\; K_V\,m f_\Delta \sum_{p} \beta_p\,e^{-j\frac{2\pi}{N}k(\ell_p-\ell_0)}.
\end{equation}
For a single path with $\ell_p{=}\ell_0$, this reduces to a flat gain $H^{\rm eff}_{k,k}=K_V\,m f_\Delta$ and zero \ac{ICI}. Under strong time selectivity (large $\delta\beta_p[n]$), off-diagonal terms grow and $\mathbf{H}_{\rm eff}$ becomes banded, calling for equalization or per-subblock processing \cite{hernando2025channel}.

\subsection{Remarks on Validity and Practical Implementation}
\label{subsec:practical_notes}
The discriminator output approximation in \eqref{eq:demod_sum} holds when the per-symbol phase dispersion is limited and the weights $\beta_p[n]$ vary slowly. Specifically, if the complex path phasors satisfy $\max_{p,p'} |\angle a_p - \angle a_{p'}| \le \pi/2$ and $|\dot\beta_p|\!\ll\! 1/T_{\mathrm{sym}}$, the weights remain non-negative and the ICI term $H^{(\mathrm{ICI})}$ remains negligible.

Regarding synchronization, the block de-meaning operation in \eqref{eq:meanrm} effectively removes the constant component of the CFO within each block. For independent samples, the residual variance is scaled by $(1-1/N)$; however, under significant Doppler drift, a small residual bias $O(\sum_p \beta_p \dot{\nu}_p T_{sym})$ may persist, which is handled by the robust phase-differencing estimator in the receiver stage.
\begin{figure}[t]
    \centering
    \includegraphics[width=\columnwidth]{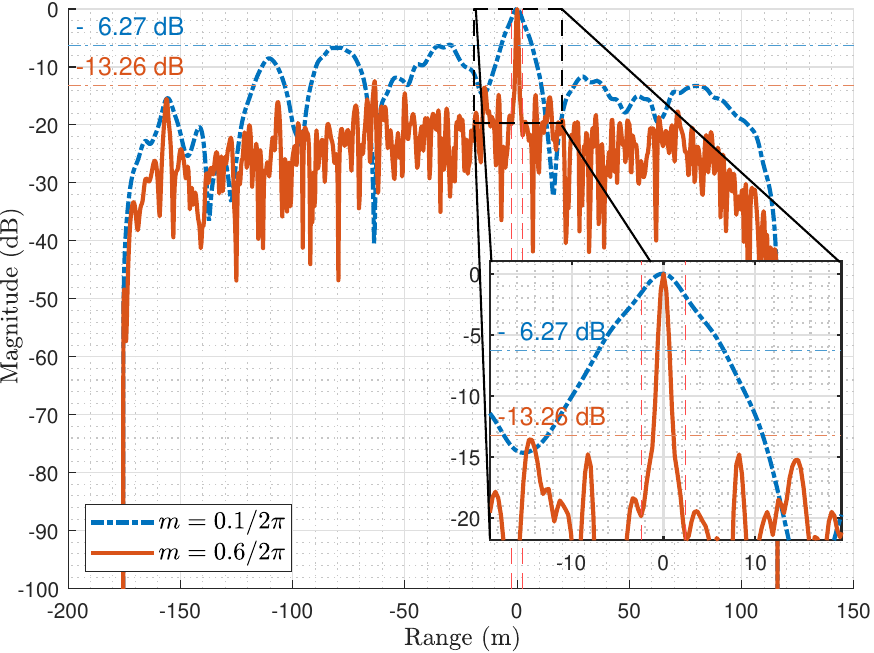}
    \caption{Effect of modulation index $m$ on the magnitude of the range-compressed output $\bar{C}[p]$ for three targets.}
    \label{fig:crosscorr_vs_m}
\end{figure}
\section{MODULATION INDEX ANALYSIS}\label{sec:mod_index}

The modulation index \(m\) serves as the principal control parameter in \ac{FM-OFDM}, determining the trade-off among spectral occupancy, demodulation \ac{SNR}, and phase-unwrapping robustness. 
Unlike \ac{CP-OFDM} or \ac{CE-OFDM}, \ac{FM-OFDM} conveys information through the instantaneous frequency deviation \cite{hernando2022frequency, hernando2025channel},
\begin{equation}
\textcolor{black}{
f[n]=\frac{m}{N_a}x[n], \qquad 
\phi[n]=\phi_0+2\pi\!\!\sum_{t\le n}\! f[t],}
\end{equation}
\textcolor{black}{
resulting in a CE signal \(s[n]=e^{j\phi[n]}\). 
Increasing \(m\) improves noise immunity but expands the occupied bandwidth and raises phase-wrap probability.}

\vspace{3pt}
\subsection{\textcolor{black}{Theoretical Impact of \(m\)}}
\textcolor{black}{
Assuming perfect phase unwrapping, the discriminator-domain \ac{SNR} scales quadratically with \(m\):}
\begin{equation}
    \text{SNR}_k^{\text{FM}} \propto 
    m^2\!\!\left(1-\cos\!\frac{2\pi k}{N}\right)\!\text{SNR}_{\text{in}},
\end{equation}

implying that the subcarrier-wise \ac{SNR} gain grows with \(m^2\).
The effective \ac{RMS} bandwidth follows a Carson-type approximation,
\begin{equation}
    B_{\text{FM}}\!\approx\! 2m f_s + \frac{N_a}{2T_s},
    \label{eq:alias}
\end{equation}

revealing a direct coupling between noise resilience and spectral expansion.

\vspace{3pt}
\subsection{\textcolor{black}{Wrapping Threshold and Statistical Bound}}
\textcolor{black}{
The instantaneous phase increment 
\(\Delta\phi[n]=2\pi m x[n]/N_a\) must stay below a limit to avoid unwrap ambiguity. 
A probabilistic bound ensures robust demodulation:}
\begin{equation}
\textcolor{black}{
P(|\Delta\phi|>\theta_{\max})\le\varepsilon, \quad
m \le \frac{\theta_{\max}N_a}{2\pi\,\text{pct}_{1-\varepsilon}(|x[n]|)},}
\end{equation}
\textcolor{black}{
where \(\theta_{\max}\!\in[1.5,2.5]\) rad and 
\(\varepsilon\!\in[10^{-4},10^{-3}]\) define the acceptable wrap margin. 
For the adopted numerology, unwrap errors typically appear for \(m\!\gtrsim\!2{-}3\).}

\vspace{3pt}
\subsection{\textcolor{black}{Effect on Sensing Performance}}
\textcolor{black}{
A higher \(m\) narrows the mainlobe of the range correlation and reduces \ac{RMSE} in range/velocity estimation up to the wrap threshold. 
Beyond this point, phase discontinuities generate sidelobes and \ac{BER} floors. 
An empirical operating range of \(0.3\!\le m\!\le\!0.7\) provides accurate sensing without spectral-mask violation.}

\subsection{Strict Bandwidth Normalization for Fair Benchmarking}
\label{subsec:fairness}
A critical methodological challenge in evaluating \ac{FM-OFDM} is that the occupied bandwidth ($B_{99}$) expands with the modulation index $m$ according to Carson's rule. Comparing high-$m$ \ac{FM-OFDM} against a standard \ac{CP-OFDM} signal with fixed subcarrier spacing is inherently unfair, as the former utilizes more spectral resources to achieve higher resolution.
\begin{figure}[t]
    \centering
    \includegraphics[scale=0.58]{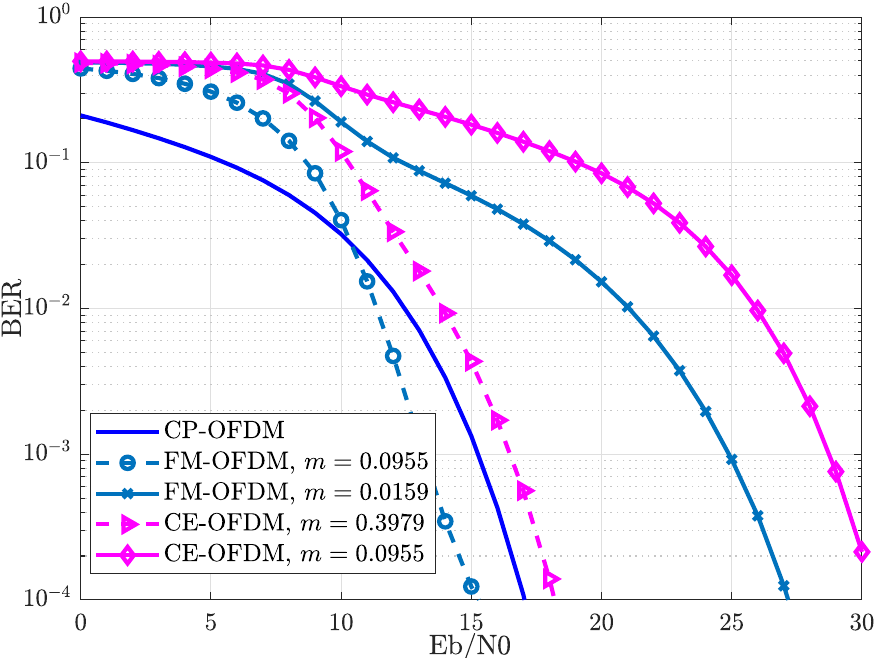}
    \caption{BER in \ac{AWGN} channel.}
    \label{fig:ber_awng}
\end{figure}
To address this, we enforce a strict bandwidth normalization constraint in all subsequent comparisons. Let $B_{\text{ref}}$ be the allocated channel bandwidth. For a given modulation index $m$, the number of active subcarriers $N_a$ in \ac{FM-OFDM} is reduced such that:
\begin{equation}
    2(m f_\Delta + B_{\text{x}}) \approx B_{\text{ref}}
\end{equation}
This ensures that \ac{FM-OFDM} ($m=0.6$) and the baselines (\ac{CP-OFDM}, \ac{CE-OFDM}) occupy the exact same spectral footprint. Consequently, any performance gain observed in Section \ref{sec:sim_results} is attributable to the waveform's structure, not unequal bandwidth usage.
%
%
%

\section{RADAR PROCESSING}
\label{sec:radar_processing}

This section presents range and velocity estimation for the proposed \ac{FM-OFDM} sensing receiver. The derivations build on Sections \ref{sec:fmofdm_io_main} and \ref{sec:mod_index}, using matched filtering in fast time and phase-based Doppler estimation in slow time.

\subsection{RECEIVED SIGNAL MODEL}
We consider $L$ point targets with ranges $\{R_\ell\}$, radial velocities $\{v_\ell\}$, and complex reflection coefficients $\{\alpha_\ell\}$. The discrete round-trip delay is $\tau_\ell \triangleq \lfloor 2 R_\ell F_s / c \rfloor$, where $F_s = 1/T_s$ is the sampling rate and $c$ is the speed of light. The Doppler frequency is $\nu_\ell \triangleq 2 v_\ell f_c / c$, where $f_c$ is the carrier frequency. Letting $u\in\{0,\ldots,U{-}1\}$ index OFDM slow time symbols and $n\in\{0,\ldots,N_{\mathrm{FFT}}{-}1\}$ index samples (where $N_{\mathrm{FFT}} = N$), the \ac{CP}-removed, complex baseband received samples for the $u$-th symbol are
\begin{equation}
    r_u[n] \;=\; \sum_{\ell=1}^{L} \alpha_\ell\, s_{\mathrm{tx}}[\,n-\tau_\ell\,]\,
    e^{j 2\pi \nu_\ell \big(nT_s + uT_{\mathrm{sym,eff}}\big)} \;+\; w_u[n],
    \label{eq:rx_signal_radar}
\end{equation}
where $s_{\mathrm{tx}}[n]$ is the \ac{CP}-free \ac{FM-OFDM} symbol, $T_{\mathrm{sym,eff}}\triangleq (N_{\mathrm{FFT}}+N_{\mathrm{CP}})T_s$ is the symbol duration, and $N_{\mathrm{CP}}\!\ge\!\tau_{\max}$.

\subsection{RANGE ESTIMATION VIA MATCHED FILTERING}
\label{subsec:range_estimation}
For symbol $u$, the fast-time \ac{MF} output at lag $p\in\mathbb{Z}$ is
\begin{equation}
    C_u[p] \;\triangleq\; \mathrm{IFFT}\!\left\{\, R_u[k]\; S_{\mathrm{tx}}^{*}[k] \,\right\},
    \label{eq:crosscorr_freq}
\end{equation}
where $R_u[k]$ and $S_{\mathrm{tx}}[k]$ are the $N_{\mathrm{FFT}}$-point \ac{DFT}s.

\subsubsection{SINGLE-TARGET RESPONSE}
For $L{=}1$, the response is
\begin{equation}
    C_u[p] = \alpha\, e^{j 2\pi \nu \big(pT_s + uT_{\mathrm{sym,eff}}\big)} A(\Delta p,\nu),
\label{eq:rnd}
\end{equation}
where $\Delta p\triangleq p-\tau$ and $A(\Delta p,\nu)$ is the ambiguity function of $s_{\mathrm{tx}}[n]$. Due to the continuous phase evolution of FM-OFDM, the mainlobe degradation $|A(0,\nu)|$ closely follows the standard sinc approximation, mitigating the losses often seen in discontinuous waveforms.

\subsubsection{RANGE MAPPING AND RESOLUTION}
Noncoherent averaging $\bar{C}[p] = \frac{1}{U}\sum_{u}|C_u[p]|$ suppresses noise. Detected peaks $\hat{\tau}_\ell$ map to range $R_\ell$ with resolution:
\begin{equation}
    \Delta R \;=\; \frac{c}{2 B_{\mathrm{eff}}}, 
    \qquad B_{\mathrm{eff}} \;\approx\; B_x + \eta\, m f_\Delta,
    \label{eq:range_res}
\end{equation}
consistent with the bandwidth analysis in Section \ref{sec:mod_index}. Fig. \ref{fig:crosscorr_vs_m} illustrates the sharpening of the mainlobe as $m$ increases.
\begin{figure}[t]
    \centering
    \includegraphics[scale=0.58]{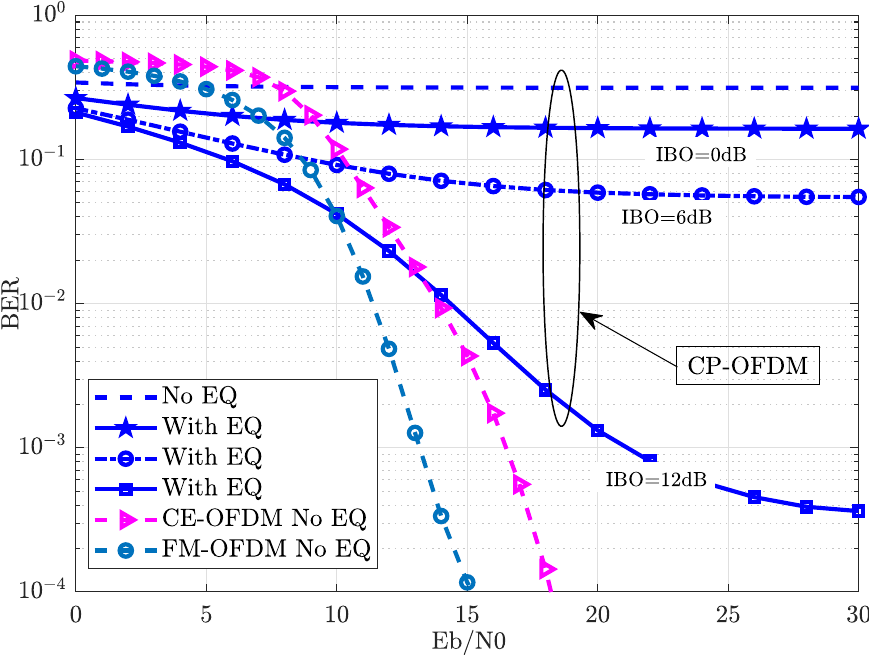}
    \caption{BER in \ac{AWGN} channel with \ac{PA}.}
    \label{fig:BER_awgn_PA}
\end{figure}
\subsection{VELOCITY ESTIMATION VIA SLOW-TIME PHASE DIFFERENCING}
\label{subsec:velocity_estimation}
After range compression, the slow-time sample at the detected delay bin $\hat{\tau}_\ell$ is
\begin{equation}
    y_\ell[u] \triangleq C_u[\hat{\tau}_\ell] \approx \alpha_\ell e^{j(2\pi\nu_\ell uT_{\mathrm{sym,eff}} + \theta_\ell[u])} + w_\ell[u].
    \label{eq:slow_time_model}
\end{equation}
The phase term $\theta_\ell[u]$ contains deterministic variations due to the FM-OFDM data sequence. A standard Doppler FFT would require explicit compensation of $\theta_\ell[u]$ (using known Tx data). To reduce complexity and improve robustness against synchronization errors, we adopt a \textbf{slow-time phase differencing} approach.

The consecutive phase increments are:
\begin{align}
    \Delta\phi_\ell[u] &= \arg\left(y_\ell[u] y_\ell^*[u-1]\right) \nonumber \\
    &\approx 2\pi\nu_\ell T_{\mathrm{sym,eff}} + \underbrace{(\theta_\ell[u] - \theta_\ell[u-1])}_{\approx 0} + \xi_\ell[u],
    \label{eq:phase_diff}
\end{align}
where the term $(\theta_\ell[u] - \theta_\ell[u-1])$ is negligible due to the slowly varying nature of the FM-OFDM phase path. The Doppler frequency is estimated by averaging the unwrapped differences:
\begin{equation}
    \hat{\nu}_\ell = \frac{1}{2\pi T_{\mathrm{sym,eff}}(U-1)} \sum_{u=1}^{U-1} \mathcal{U}(\Delta\phi_\ell[u]).
\end{equation}
The radial velocity is then $\hat{v}_\ell = \frac{\lambda}{2} \hat{\nu}_\ell$.

\subsubsection{Performance Analysis}
The estimator provides an unambiguous velocity range of $v_{\max} = \pm \lambda / (4T_{\mathrm{sym,eff}})$. The theoretical variance approximates:
\begin{equation}
    \mathrm{Var}(\hat{v}_\ell) \approx \frac{\lambda^2}{8\pi^2 T_{\mathrm{sym,eff}}^2 (U-1) \gamma_\ell},
    \label{eq:velocity_variance}
\end{equation}
where $\gamma_\ell$ is the post-compression SNR. This method avoids the structural error floors caused by the data-dependent phase $\theta_\ell[u]$ in standard FFT processing, as demonstrated in Section \ref{sec:sim_results}.
\begin{figure}[t]
    \centering
    \includegraphics[scale=0.58]{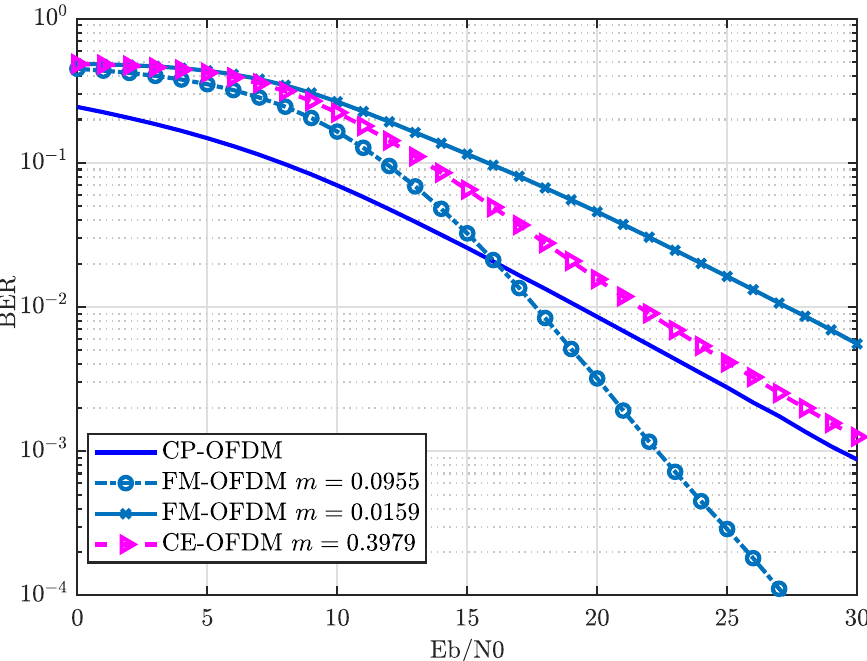}
    \caption{BER performance in a 5-tap doubly dispersive Rayleigh channel ($v_{max}= 200$km/h).}
    \label{fig:ber_dd}
\end{figure}
\section{SIMULATION RESULTS AND ANALYSIS}
\label{sec:sim_results}

In this section, a detailed performance evaluation of the proposed \ac{FM-OFDM} waveform in both communication and sensing scenarios is presented. To ensure a fair comparison between (\ac{FM-OFDM}, \ac{CE-OFDM}) and standard \ac{CP-OFDM}, we adopt the following parameters; an FFT size of $N_{\mathrm{FFT}}=512$, a subcarrier spacing of $15$ kHz. The carrier frequency is set to $f_c = 77$ GHz for automotive scenarios and $2.4$ GHz for Wi-Fi sensing cases. All waveforms utilize 64-\ac{QAM} modulation. The choice of parameters is to compare between the waveforms under the same umbrella. 

For the nonlinear \ac{PA} analysis (Fig. \ref{fig:BER_awgn_PA}), we use Saleh model for the \ac{PA} \cite{o2009new}. The Input Back-Off (IBO)$\text{IBO}=0$ dB corresponds to fully saturated operation. The "With EQ" curves denote performance after standard zero forcing frequency domain equalization, which attempts to compensate for both channel effects and warping induced by the \ac{PA} nonlinearity.

\begin{figure}[t]
    \centering
    \includegraphics[scale=0.58]{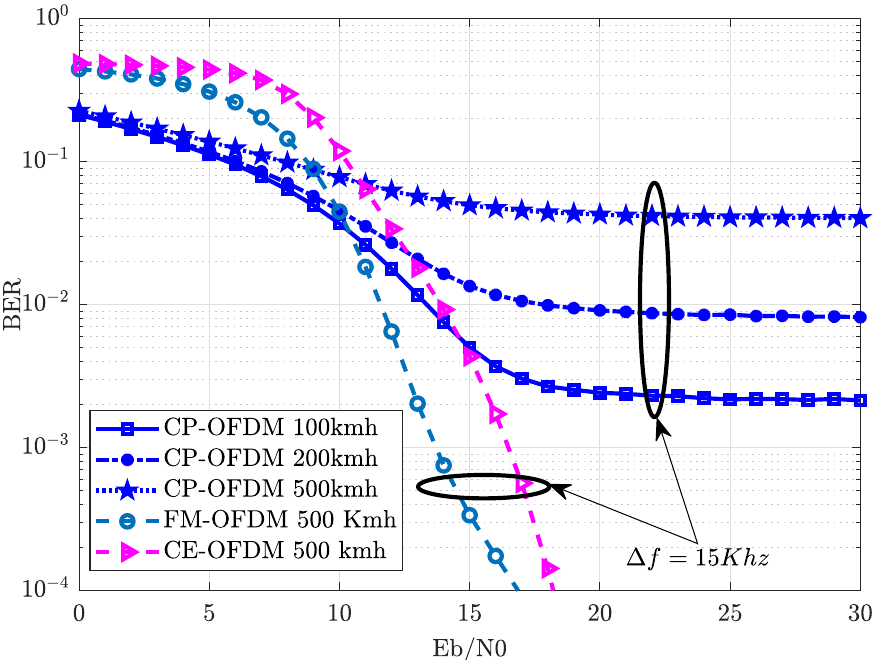}
    \caption{BER performance in single-tap time-selective channel vs. speed.}
    \label{fig:ber_stap_channel}
\end{figure}

\begin{figure}[t]
    \centering
    \includegraphics[scale=0.58]{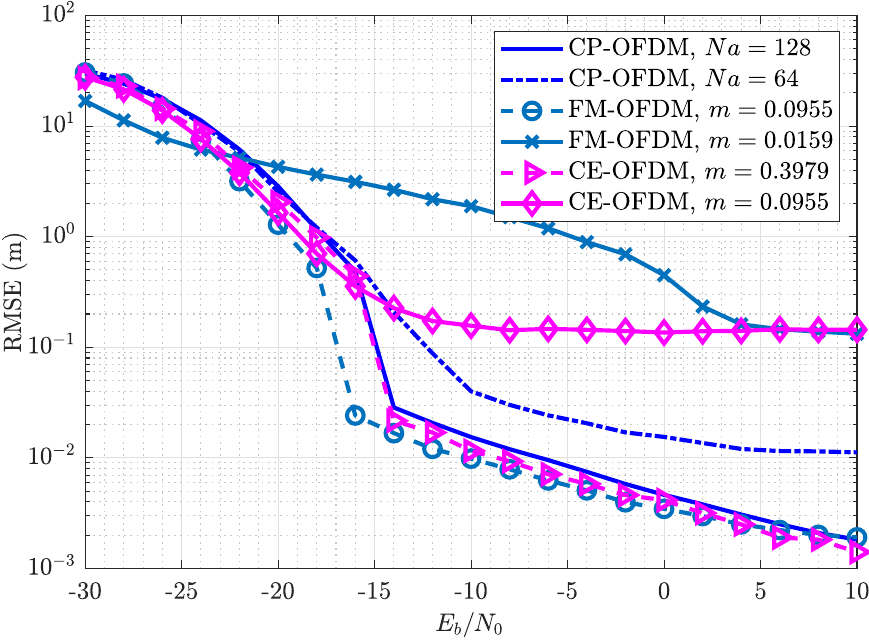}
    \caption{Range RMSE vs. SNR.}
    \label{fig:rmse_rng}
\end{figure}

A critical aspect of this study is the strict normalization of the occupied bandwidth ($B_{99}$).
For the \ac{BER} analyses (Figs. \ref{fig:ber_awng} \ref{fig:ber_stap_channel}), The active subcarriers is fixed to $N_a=64$ for all waveforms to evaluate fundamental robustness under identical subcarrier spacing $\Delta f$.    Sensing analyses are shown in (Figs. \ref{fig:rmse_rng} \ref{fig:rmse_velo}). To match their physical bandwidth usage, the benchmark \ac{CP-OFDM} is allocated $N_a=128$ active subcarriers. This ensures that any resolution gain observed in \ac{FM-OFDM} is due to the waveform structure, not simply utilizing more spectrum than the baseline.

The receiver processing differs by waveform to reflect optimal implementations. For \ac{CP-OFDM}, we employ the conventional 2D-\ac{FFT} periodogram based estimator \cite{sturm2011waveform}. For \ac{FM-OFDM} and \ac{CE-OFDM}, we utilize the proposed method described in Section \ref{sec:radar_processing}.
\subsection{BER PERFORMANCE}
\label{subsec:awgn_pa}
Figure \ref{fig:ber_awng} presents the \ac{BER} performance in an \ac{AWGN} channel. Under ideal linear conditions, the performance of the wavefomrms is consistent with prior work \cite{hernando2022frequency}. \ac{FM-OFDM} and \ac{CE-OFDM} achieve error rates comparable to \ac{CP-OFDM} without requiring power back off. These results validate the fundamental communication performance of \ac{FM-OFDM} and confirm its suitability as a dual function waveform.

Figure \ref{fig:BER_awgn_PA} demonstrates the advantage of constant envelope waveforms under nonlinear \ac{PA} distortion. \ac{FM-OFDM} and \ac{CE-OFDM} maintain nearly identical \ac{BER} performance to the linear case, while \ac{CP-OFDM} suffers significant degradation due to its high \ac{PAPR} which is similar to results in \cite{hadani2018otfs}. This result directly validates the theoretical hardware benefits of \ac{FM-OFDM}'s CE property, ensuring that ISAC transceivers can maximize detection range by utilizing the full saturated power of the \ac{PA}.

Figure \ref{fig:ber_dd} presents the BER performance in a doubly dispersive five-tap Rayleigh channel, validating the analytical framework developed in Section \ref{sec:fmofdm_io_main}. \ac{FM-OFDM} attains the lowest \ac{BER} across the entire \ac{SNR} range, particularly for $m=0.6/(2\pi)$, confirming the Doppler resilience and continuous-phase advantage predicted by the analytical model.

The performance improvement with increasing modulation index $m$ aligns with the theory. As $m$ increases, the instantaneous frequency deviation enlarges. Consistent with the analysis in Section \ref{sec:mod_index} (Eq. \ref{eq:alias}), this expands the effective bandwidth $B_{\mathrm{eff}}$, thereby sharpening the correlation mainlobe and improving resolution.

Figure \ref{fig:ber_stap_channel} further demonstrates \ac{FM-OFDM}'s robustness under high mobility conditions. As mobility increases from 300 km/h to 800 km/h, \ac{CP-OFDM} suffers rapid performance degradation due to Doppler induced loss of subcarrier orthogonality. \ac{CE-OFDM} exhibits an error floor caused by phase unwrapping failures \cite{chung1999constant}. FM-\ac{OFDM} maintains the lowest \ac{BER} due to its continuous phase evolution, which preserves slow time coherence and limits \ac{ICI}.

\subsection{RANGE AND VELOCITY RMSE}
\label{subsec:rmse}
Simulation were done over 3 targets with fixed different ranges and velocities. The goal was to compare the performance under same conditions given a set of \ac{SNR} values. 
As shown in Fig. \ref{fig:rmse_rng}, \ac{FM-OFDM} with $m \approx 0.6/(2\pi)$ achieves an RMSE comparable to \ac{CP-OFDM} ($N_a=128$) benchmark, settling at the theoretical bound. This confirms that with sufficient modulation index, FM-OFDM recovers the full range resolution of the equivalent-bandwidth OFDM signal. In contrast, the low-index case ($m \approx 0.1/(2\pi)$) hits an error floor, confirming that bandwidth expansion is the dominant factor for delay resolution.

\begin{figure}[t]
    \centering
    \includegraphics[scale=0.58]{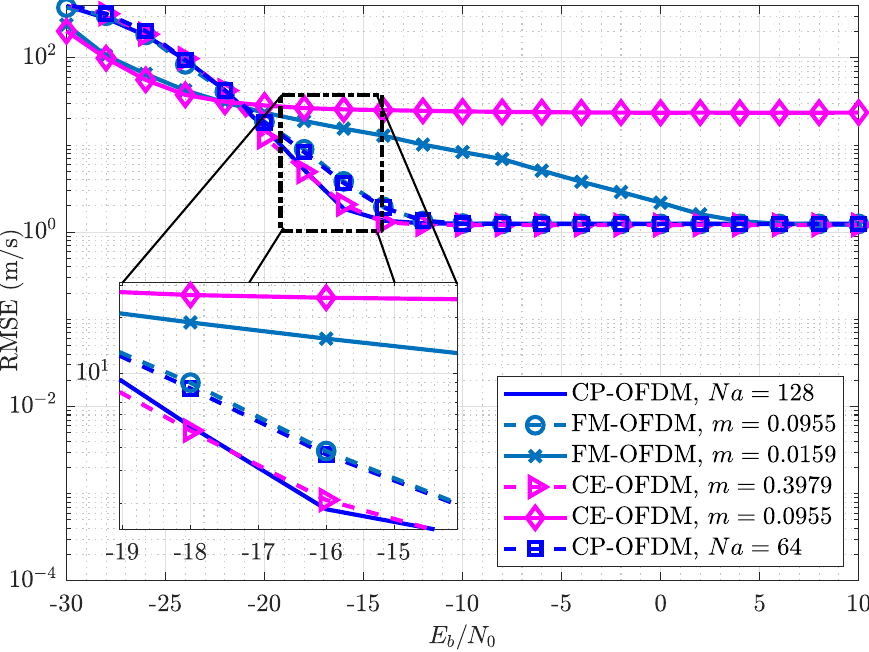}
    \caption{Velocity RMSE vs. SNR.}
    \label{fig:rmse_velo}
\end{figure}

Fig. \ref{fig:rmse_velo} shows advantage of \ac{FM-OFDM} over other schemes. \ac{CE-OFDM} waveforms (pink curves) hit an error floor at $\approx 10^{-1}$ m/s, unable to improve with SNR due to phase-unwrapping errors. \ac{FM-OFD} ($m \approx 0.6/(2\pi)$) closely tracks the performance of the ideal CP-OFDM ($N_a=128$) benchmark, descending to $\approx 10^{-3}$ m/s at high SNR.
This result proves that \ac{FM-OFDM} achieves comparable performance with \ac{OFDM} with a CE waveform that maximizes \ac{PA} efficiency.

\subsection{RANGE-DOPPLER MAP VISUALIZATION}
\label{subsec:rdm}
\begin{figure*}[t]
    \centering
    \begin{subfigure}{0.31\textwidth}
        \includegraphics[width=\linewidth]{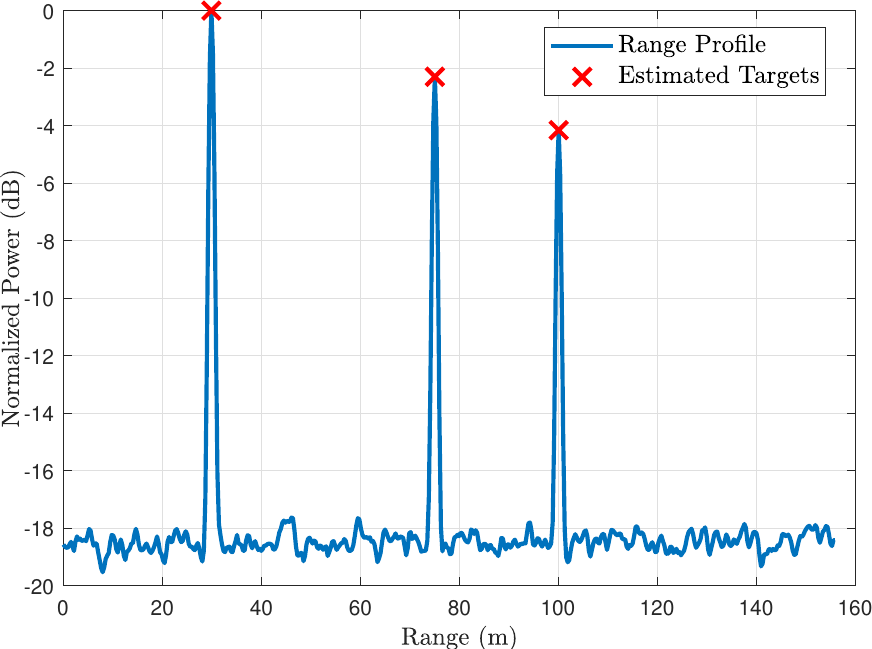}
        \caption{Range profile ($m=0.6/2\pi$).}
    \end{subfigure}\hfill
    \begin{subfigure}{0.31\textwidth}
        \includegraphics[width=\linewidth]{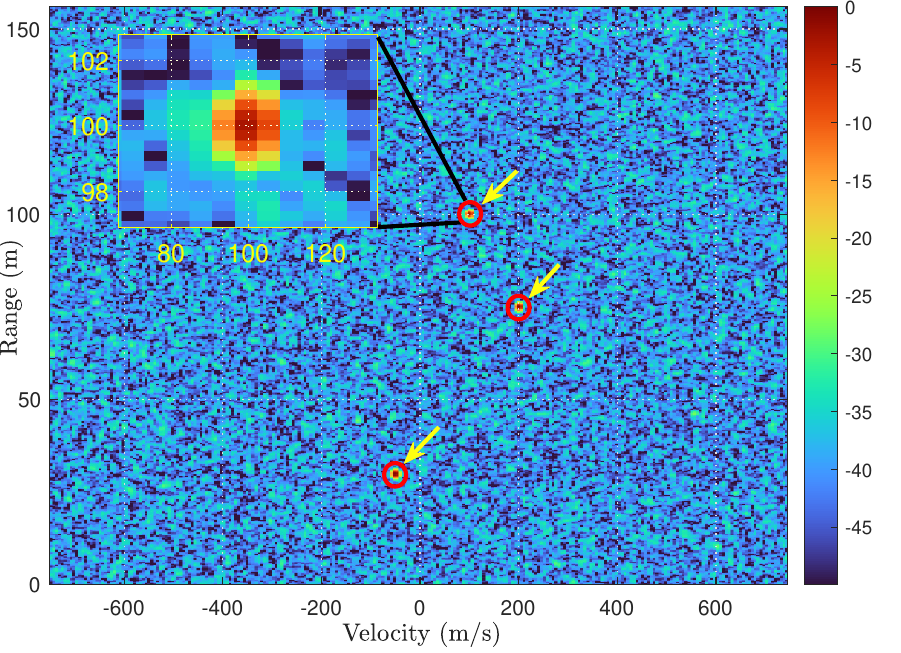}
        \caption{2D Map ($m=0.6/2\pi$).}
    \end{subfigure}\hfill
    \begin{subfigure}{0.31\textwidth}
        \includegraphics[width=\linewidth]{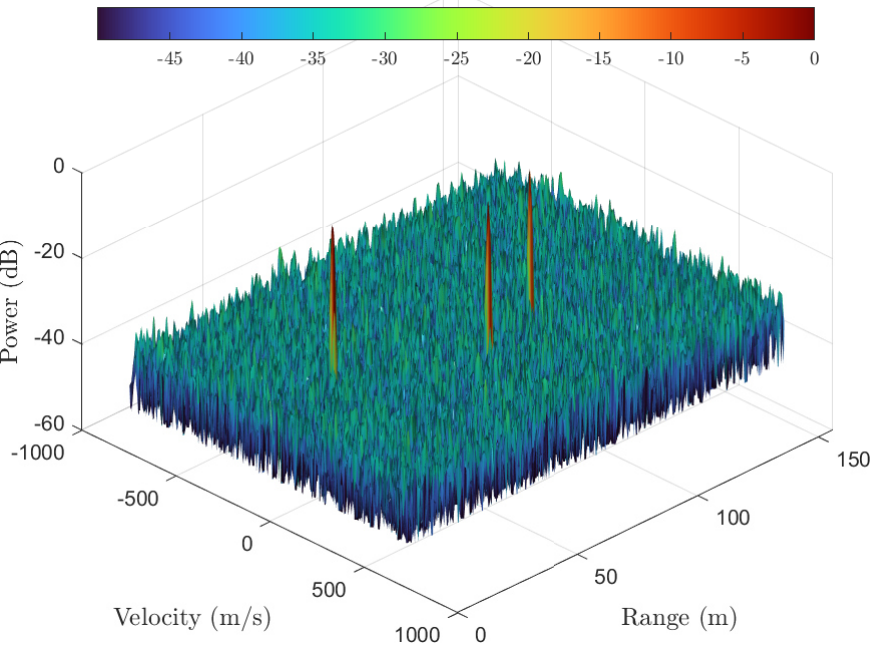}
        \caption{3D Map ($m=0.6/2\pi$).}
    \end{subfigure}
    \smallskip
    \begin{subfigure}{0.31\textwidth}
        \includegraphics[width=\linewidth]{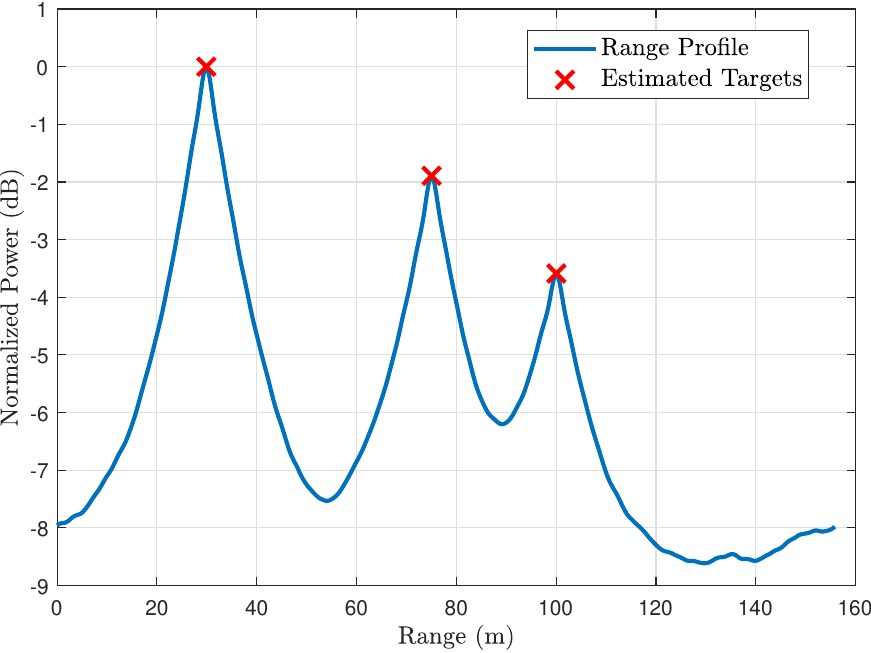}
        \caption{Range profile ($m=0.1/2\pi$).}
    \end{subfigure}\hfill
    \begin{subfigure}{0.31\textwidth}
        \includegraphics[width=\linewidth]{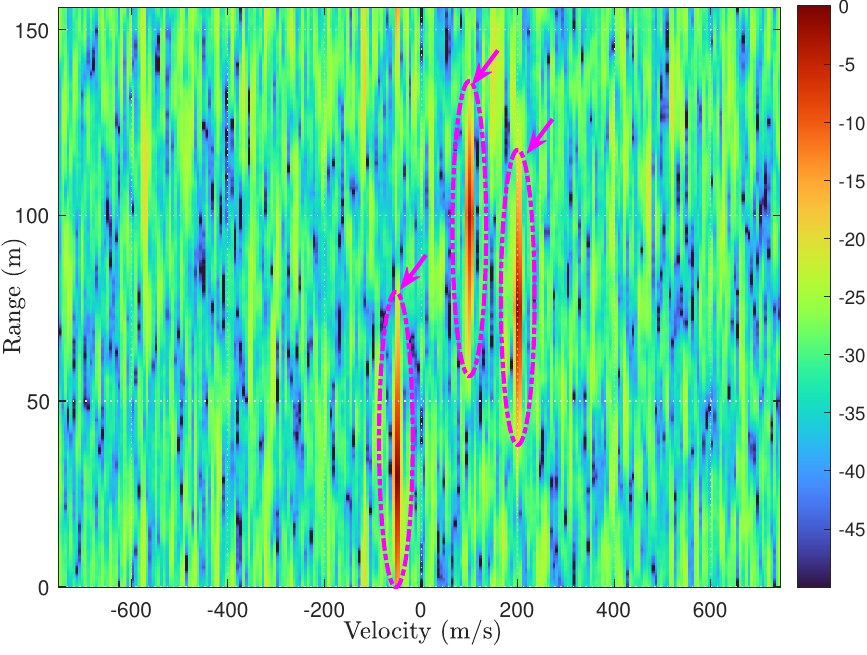}
        \caption{2D Map ($m=0.1/2\pi$).}
    \end{subfigure}\hfill
    \begin{subfigure}{0.31\textwidth}
        \includegraphics[width=\linewidth]{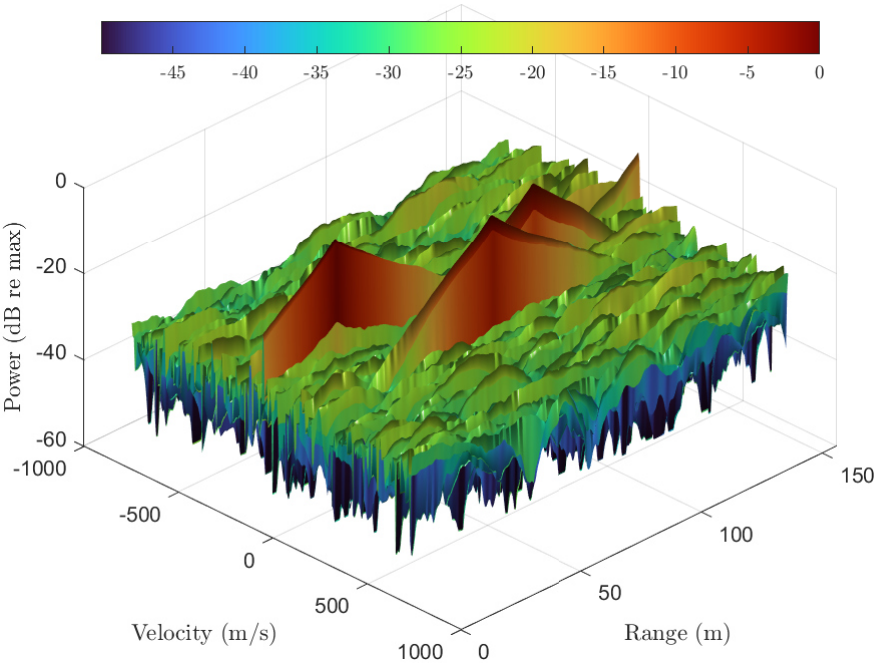}
        \caption{3D Map ($m=0.1/2\pi$).}
    \end{subfigure}
    \caption{Range profiles and Range–Doppler maps at SNR = 0 dB for modulation indices $m=0.6/2\pi$ and $m=0.1/2\pi$.}
    \label{fig:range_profil_rdms}
\end{figure*}

Figure \ref{fig:range_profil_rdms} visualizes the sensing capability at 0 dB SNR. For $m=0.6/(2\pi)$, the range profile exhibits sharp, well-separated peaks with low sidelobes, confirming the high resolution predicted by Eq. \eqref{eq:range_res}. The corresponding Range-Doppler maps display clear target clusters with limited noise spreading, effectively acting as an empirical Ambiguity Function that validates the waveform's "thumbtack" resolution properties. Conversely, the $m=0.1/(2\pi)$ case shows broadened mainlobes and overlapping returns, validating the bandwidth-resolution trade-off derived in Section \ref{sec:mod_index}.

\subsection{Practical Considerations}
Increasing the modulation index $m$ enhances sensing performance but increases the effective bandwidth $B_{\mathrm{eff}}$. While $m=0.9/(2\pi)$ yields optimal results in high-SNR scenarios, previous studies \cite{hernando2022frequency} recommend $m \approx 0.6/(2\pi)$ as a robust compromise for communication-centric applications. These findings support adaptive modulation strategies that dynamically balance sensing resolution and spectral containment based on real-time requirements.

\section{CONCLUSION}
\label{sec:conclusion}
This paper presented a comprehensive investigation of \ac{FM-OFDM} as a unified waveform for \ac{6G} ISAC. Through analytical framework and normalized bandwidth benchmarking, FM-OFDM effectively addresses the \ac{PAPR} limitations of conventional \ac{OFDM} while providing comparable sensing resolution.
Results show that \ac{FM-OFDM} enables fully saturated \ac{PA} operation without the \ac{BER} penalty seen in \ac{CP-OFDM}, directly translating to increased sensing range and energy efficiency. Furthermore, the proposed discriminator domain sensing receiver was shown to robustly estimate velocity in high mobility scenarios. By validating these performance gains under strict equal-bandwidth constraints, we established \ac{FM-OFDM} as a power-efficient alternative and a high-performance candidate for next-generation \ac{ISAC} systems.

\appendices
\section{Sum of Paths Demods to a Weighted Sum of Instantaneous Frequencies}
\label{app:weights}

\subsection{CONTINUOUS TIME-IDENTITY}
Let $z(t)=\sum_{p=0}^{P-1} e^{j\theta_p(t)}$ be the sum of unit-magnitude phasors with differentiable phases $\theta_p(t)$. The instantaneous angular frequency of $z(t)$ is
\begin{equation}
    \dot\Theta(t) \triangleq \frac{d}{dt}\arg z(t) \;=\;
    \frac{\operatorname{Im}\!\big\{z^*(t)\,\dot z(t)\big\}}{|z(t)|^2}. \label{eq:instfreq_id}
\end{equation}
Compute $z^*\dot z = j\sum_{p,q}\dot\theta_p e^{j(\theta_p-\theta_q)}$, so
\begin{align}
    \operatorname{Im}\!\{z^*\dot z\}
          & = \sum_{p,q}\dot\theta_p \cos(\theta_p-\theta_q)
    = \sum_p \dot\theta_p \underbrace{\sum_q \cos(\theta_p-\theta_q)}_{S_p}, \\
    |z|^2 & = \sum_{u,v}\cos(\theta_u-\theta_v) \triangleq S.
\end{align}
Thus
\begin{equation}
    \dot\Theta(t)=\sum_{p=0}^{P-1}\beta_p(t)\,\dot\theta_p(t),\qquad
    \beta_p(t)\triangleq \frac{S_p}{S}, \quad \sum_p \beta_p(t)=1. \label{eq:beta_def}
\end{equation}
For \ac{FM-OFDM}, each path phase is
\begin{align}
    \theta_p(t)      & = \arg a_p + 2\pi \nu_p t + \phi(t-\tau_p),    \\
    \dot\theta_p(t) & = 2\pi\big(\nu_p + m f_\Delta \,x(t-\tau_p)\,\big)
\end{align}
where $x(t)$ denotes the continuous-time counterpart of the discrete-time sequence $x[n]$ defined in Section \ref{sec:fmofdm_io_main}. Hence \eqref{eq:beta_def} yields
\begin{equation}
    \dot\Theta(t)=2\pi\sum_p \beta_p(t)\big(\nu_p + m f_\Delta x(t-\tau_p)\big).
\end{equation}

\subsection{DISCRETE TIME-DISCRIMINATOR}
Let $z[n]=r[n]/|r[n]|$ be the limited signal. The standard one-sample phase-difference estimator
\[
    \widehat f[n] = \frac{1}{2\pi T_s}\angle\!\big(z[n]z^*[n-1]\big)
\]
is a first-order finite-difference approximation of $\dot\Theta(t)/(2\pi)$ provided the per-sample phase increment is $<\pi$. Replacing $t$ by $nT_s$ and using the circular block model, we obtain
\begin{equation}
    \widehat f[n]\approx \sum_p \beta_p[n]\big(\nu_p + m f_\Delta x[n-\ell_p]\big).
\end{equation}
Multiplying by $K_V$ gives \eqref{eq:demod_sum} in the main text.

\bibliographystyle{IEEEtran}
\bibliography{IEEEfull}

@article{osorio2025rise,
  title={The Rise of Networked {ISAC}: Emerging Aspects and Challenges},
  author={Osorio, Diana PM and Barua, Bidushi and Besser, Karl-Ludwig and Blue, Henry and Dass, Prajnamaya and Porambage, Pawani},
  journal={IEEE Open Journal of the Communications Society},
  year={2025},
  publisher={IEEE}
}

@article{chen2022antenna,
  title={Antenna/propagation domain self-interference cancellation ({SIC}) for in-band full-duplex wireless communication systems},
  author={Chen, Yuenian and Ding, Can and Jia, Yongtao and Liu, Ying},
  journal={Sensors},
  volume={22},
  number={5},
  pages={1699},
  year={2022},
  publisher={MDPI}
}

@article{jeong2025interference,
  title={Interference Analysis and Successive Interference Cancellation for Multistatic {OFDM}-based {ISAC} Systems},
  author={Jeong, Taewon and Giroto, Lucas and Erdem, Umut Utku and Karle, Christian and Choi, Jiyeon and Zwick, Thomas and Nuss, Benjamin},
  journal={arXiv preprint arXiv:2507.20942},
  year={2025}
}

@article{zhuo2022multibeam,
  title={Multibeam joint communication and radar sensing: Beamforming design and interference cancellation},
  author={Zhuo, Yinxiao and Sha, Ziyuan and Wang, Zhaocheng},
  journal={IEEE Communications Letters},
  volume={26},
  number={8},
  pages={1888--1892},
  year={2022},
  publisher={IEEE}
}

@article{hadani2018otfs,
  title={{OTFS}: A new generation of modulation addressing the challenges of 5{G}},
  author={Hadani, Ronny and Monk, Anton},
  journal={arXiv preprint arXiv:1802.02623},
  year={2018}
}

@inproceedings{wang2022triangular,
  title={Triangular {FM-OFDM} waveform design for integrated sensing and communication},
  author={Wang, Yuan and Wei, Zhiqing and Zhou, Wei and Han, Kaifeng and Feng, Zhiyong},
  booktitle={2022 IEEE International Conference on Communications Workshops (ICC Workshops)},
  pages={515--519},
  year={2022},
  organization={IEEE}
}

@article{GonzalezPrelcic_ProcIEEE_2024,
  author  = {Nuria Gonz{\'a}lez-Prelcic and Musa Furkan Keskin and Ossi Kaltiokallio and Mikko Valkama and Davide Dardari and Xiao Shen and Yuan Shen and Murat Bayraktar and Henk Wymeersch},
  title   = {The Integrated Sensing and Communication Revolution for 6{G}: Vision, Techniques, and Applications},
  journal = {Proceedings of the IEEE},
  year    = {2024},
  doi     = {10.1109/JPROC.2024.XXXXX}, 
  note    = {Early access / July 2024 issue; also available as arXiv:2405.01816}
}

@article{Oliveira_PN_OFDM_ISAC_2024,
  author  = {Guilherme F. Oliveira and Pedro M. Duarte and Edson F. de Souza and Ricardo A. A. de Souza and Moab A. R. Perrel},
  title   = {On the Sensing Performance of {OFDM}-Based {ISAC} Under the Influence of Oscillator Phase Noise},
  journal = {arXiv preprint},
  year    = {2024},
  eprint  = {2410.13336},
  url     = {https://arxiv.org/abs/2410.13336}
}

@article{Chen_Golay_CE_OFDM_ISAC_2024,
  author  = {Xinyu Chen and Bin Rao and Dan Song and Wei Wang and Xiaohai Zou},
  title   = {Golay Complementary Sequence and Constant Envelope Orthogonal Frequency-Division Multiplexing-Based for Integrated Sensing and Communication with Mutual Information Analysis},
  journal = {IET Radar, Sonar \& Navigation},
  year    = {2024},
  volume  = {18},
  number  = {10},
  pages   = {1848--1858},
  doi     = {10.1049/rsn2.12622}
}

@article{hernando2025channel,
  title={Channel Estimation and Equalization of Zero-Padded Waveforms in Doubly-Dispersive Channels},
  author={Hernando, Javier Lorca and M{\'e}ndez-Monsanto, Lianet and Armada, Ana Garc{\'\i}a},
  journal={IEEE Transactions on Communications},
  year={2025},
  publisher={IEEE}
}

@misc{5GAmericas2024_3GPPTrends,
    author  = {5{G} Americas},
    title   = {{3GPP Technology Trends - White Paper 2}},
    year    = {2024},
    note    = {Accessed online: \url{https://www.5gamericas.org/wp-content/uploads/2024/01/3GPP-Technology-Trends-WP.pdf}},
    urldate = {2025-05-07}
}

@techreport{ITU_M2160_2023,
  author = {{ITU-R}},
  title   = {Framework and overall objectives of {IMT} for 2030 and beyond ({IMT}-2030)},
  institution = {ITU-R Recommendation M.2160-0},
  year    = {2023},
  url     = {https://www.itu.int/dms_pubrec/itu-r/rec/m/R-REC-M.2160-0-202311-I!!PDF-E.pdf}
}

@article{liu2025uncovering,
  title={Uncovering the iceberg in the sea: Fundamentals of pulse shaping and modulation design for random {ISAC} signals},
  author={Liu, Fan and Xiong, Yifeng and Lu, Shihang and Li, Shuangyang and Yuan, Weijie and Masouros, Christos and Jin, Shi and Caire, Giuseppe},
  journal={IEEE Transactions on Signal Processing},
  year={2025},
  publisher={IEEE}
}

@misc{3GPP_RAN_R19_ISAC,
  author = {{3GPP}},
  title = {{RAN} Rel-19 Status (includes {ISAC} channel model study item)},
  howpublished = {\url{https://www.3gpp.org/specifications/82-release-19}},
  year  = {2025},
  note  = {Accessed Jun. 16, 2025}
}

@article{liao2025pulse,
  title={Pulse shaping for random {ISAC} signals: The ambiguity function between symbols matters},
  author={Liao, Zihan and Liu, Fan and Li, Shuangyang and Xiong, Yifeng and Yuan, Weijie and Masouros, Christos and Lops, Marco},
  journal={IEEE Transactions on Wireless Communications},
  year={2025},
  publisher={IEEE}
}

@article{o2009new,
  title={New modified Saleh models for memoryless nonlinear power amplifier behavioural modelling},
  author={O'droma, Mairtin and Meza, Serban and Lei, Yiming},
  journal={IEEE Communications Letters},
  volume={13},
  number={6},
  pages={399--401},
  year={2009},
  publisher={IEEE}
}

@techreport{ETSI_GR_ISAC_001,
  author = {{ETSI ISAC ISG}},
  title = {GR {ISAC} 001: Use Cases, Requirements and System Aspects for Integrated Sensing and Communications},
  institution = {ETSI ISG ISAC},
  year  = {2024},
  url   = {https://www.etsi.org/committee/1968-isac}
}

@ARTICLE{10685511,
  author={Du, Zhen and Liu, Fan and Xiong, Yifeng and Han, Tony Xiao and Eldar, Yonina C. and Jin, Shi},
  journal={IEEE Transactions on Signal Processing}, 
  title={Reshaping the {ISAC} Tradeoff Under {OFDM} Signaling: A Probabilistic Constellation Shaping Approach}, 
  year={2024},
  volume={72},
  number={},
  pages={4782-4797},
  doi={10.1109/TSP.2024.3465499}}

@misc{IEEE80211bf_2025,
  author = {},
  title   = {{IEEE} 802.11bf—{WLAN} Sensing Amendment (Approved 2025)},
  howpublished = {\url{https://standards.ieee.org/ieee/802.11bf/11077/}},
  year    = {2025},
  note    = {Accessed Jun. 21, 2025}
}

@article{sturm2011waveform,
  title={Waveform design and signal processing aspects for fusion of wireless communications and radar sensing},
  author={Sturm, Christian and Wiesbeck, Werner},
  journal={Proceedings of the IEEE},
  volume={99},
  number={7},
  pages={1236--1259},
  year={2011},
  publisher={IEEE}
}

@inproceedings{chung1999constant,
  title={Constant-envelope orthogonal frequency division multiplexing modulation},
  author={Chung, Char-Dir and Cho, Shih-Ming},
  booktitle={Fifth Asia-Pacific Conference on... and Fourth Optoelectronics and Communications Conference on Communications,},
  volume={1},
  pages={629--632},
  year={1999},
  organization={IEEE}
}

@book{haykin2008communication,
  title={Communication systems},
  author={Haykin, Simon},
  year={2008},
  publisher={John Wiley \& Sons}
}

@article{loughlin2002comments,
  title={Comments on the interpretation of instantaneous frequency},
  author={Loughlin, Patrick J and Tacer, Berkant},
  journal={IEEE Signal Processing Letters},
  volume={4},
  number={5},
  pages={123--125},
  year={1997},
  publisher={IEEE}
}

@book{flandrin1998time,
  title={Time-frequency/time-scale analysis},
  author={Flandrin, Patrick},
  volume={10},
  year={1998},
  publisher={Academic press}
}

@article{boashash2002estimating,
  title={Estimating and interpreting the instantaneous frequency of a signal. I. Fundamentals},
  author={Boashash, Boualem},
  journal={Proceedings of the IEEE},
  volume={80},
  number={4},
  pages={520--538},
  year={1992},
  publisher={IEEE}
}

@ARTICLE{9924202,
  author={Zhou, Wenxing and Zhang, Ruoyu and Chen, Guangyi and Wu, Wen},
  journal={IEEE Open Journal of the Communications Society}, 
  title={Integrated Sensing and Communication Waveform Design: A Survey}, 
  year={2022},
  volume={3},
  number={},
  pages={1930-1949},
  doi={10.1109/OJCOMS.2022.3215683}}

@ARTICLE{9144301,
  author={Chowdhury, Mostafa Zaman and Shahjalal, Md. and Ahmed, Shakil and Jang, Yeong Min},
  journal={IEEE Open Journal of the Communications Society}, 
  title={6G Wireless Communication Systems: Applications, Requirements, Technologies, Challenges, and Research Directions}, 
  year={2020},
  volume={1},
  number={},
  pages={957-975},
  doi={10.1109/OJCOMS.2020.3010270}}

@article{fang2022joint,
    title     = {Joint communication and sensing toward 6{G}: Models and potential of using {MIMO}},
    author    = {Fang, Xinran and Feng, Wei and Chen, Yunfei and Ge, Ning and Zhang, Yan},
    journal   = {IEEE Internet of Things Journal},
    volume    = {10},
    number    = {5},
    pages     = {4093--4116},
    year      = {2022},
    publisher = {IEEE}
}

@inproceedings{felton2023gradient,
    title        = {Gradient-descent based optimization of constant envelope {OFDM} waveforms},
    author       = {Felton, David G and Hague, David A},
    booktitle    = {2023 IEEE Radar Conference (RadarConf23)},
    pages        = {1--6},
    year         = {2023},
    organization = {IEEE}
}

@article{hernando2022frequency,
  title={Frequency-modulated {OFDM}: A new waveform for high-mobility wireless communications},
  author={Hernando, Javier Lorca and Armada, Ana Garc{\'\i}a},
  journal={IEEE Transactions on Communications},
  volume={71},
  number={1},
  pages={540--552},
  year={2022},
  publisher={IEEE}
}

@article{liu2022integrated,
    title     = {Integrated sensing and communications: Toward dual-functional wireless networks for 6{G} and beyond},
    author    = {Liu, Fan and Cui, Yuanhao and Masouros, Christos and Xu, Jie and Han, Tony Xiao and Eldar, Yonina C and Buzzi, Stefano},
    journal   = {IEEE journal on selected areas in communications},
    volume    = {40},
    number    = {6},
    pages     = {1728--1767},
    year      = {2022},
    publisher = {IEEE}
}

@article{liu2022survey,
  title={A survey on fundamental limits of integrated sensing and communication},
  author={Liu, An and Huang, Zhe and Li, Min and Wan, Yubo and Li, Wenrui and Han, Tony Xiao and Liu, Chenchen and Du, Rui and Tan, Danny Kai Pin and Lu, Jianmin and others},
  journal={IEEE Communications Surveys \& Tutorials},
  volume={24},
  number={2},
  pages={994--1034},
  year={2022},
  publisher={IEEE}
}

@article{liu2024next,
    title     = {Next-generation multiple access for integrated sensing and communications},
    author    = {Liu, Yaxi and Huang, Tianyao and Liu, Fan and Ma, Dingyou and Huangfu, Wei and Eldar, Yonina C},
    journal   = {Proceedings of the IEEE},
    year      = {2024},
    publisher = {IEEE}
}

@article{wei2023integrated,
    title     = {Integrated sensing and communication signals toward {5G-A} and {6G}: A survey},
    author    = {Wei, Zhiqing and Qu, Hanyang and Wang, Yuan and Yuan, Xin and Wu, Huici and Du, Ying and Han, Kaifeng and Zhang, Ning and Feng, Zhiyong},
    journal   = {IEEE Internet of Things Journal},
    volume    = {10},
    number    = {13},
    pages     = {11068--11092},
    year      = {2023},
    publisher = {IEEE}
}

@article{thompson2008constant,
    title={Constant envelope OFDM phase modulation},
    author={Thompson, Paul},
    journal={Electronics Letters},
    volume={44},
    number={3},
    pages={201--202},
    year={2008},
    publisher={IET}
}

\end{document}